\documentclass{iau} 
\usepackage{graphicx}

\title[Contribution of globular clusters to the halo] %% give here short title %%
{Globular clusters and their contribution \\ to the formation of the Galactic
halo}

\author[Eugenio Carretta]   %% give here short author list %%
{Eugenio Carretta$^1$}
\affiliation{$^1$INAF-Osservatorio Astronomico di Bologna, via Ranzani 1,
I-40127 Bologna, Italy \\ email: {\tt eugenio.carretta@oabo.inaf.it}}

\pubyear{2015}
\volume{317}  %% insert here IAU Symposium No.
\jname{The General Assembly of the Galaxy halos: Structure, origin, and
evolution}
\editors{A.Bragaglia, M.Arnaboldi, M. Rejkuba, \& D. Romano, eds.}
\begin{document}

\maketitle

\begin{abstract}
This is a ``biased" review because I will show recent evidence on the
contribution of globular clusters (GCs) to the halo of our Galaxy seen 
through the lens of the new paradigm of multiple populations in GCs. 
I will show a few examples where the chemistry of multiple populations helps to
answer hot questions including whether and how much GCs did contribute to the
halo population, if we have evidence of the GCs-halo link, what are the
strengths and weak points concerning this contribution.
\end{abstract}

\firstsection % if your document starts with a section,
\section{Introduction}

We obviously know that GCs {\textit do contribute} to the Galactic halo, 
there are about
100 ``smoking guns", i.e. the GCs currently observed in the Milky Way (MW) halo. Furthermore,
large surveys like SDSS, Pan-STARRS, Dark Energy Survey, public ESO surveys such
as VVV and ATLAS routinely discover
new systems in the grey zone between
globular clusters and dwarf galaxies, sometimes with controversial
classification, see e.g. the case of Crater/Laevens1 considered to be a dwarf galaxy
(\cite[Bonifacio \etal\ 2015]{bon15}) or a GC (\cite[Kirby \etal\ 2015a]{kir15a}).

Mass loss is expected for GCs in particular at early phases, during the violent
relaxation following gas expulsion (see e.g. \cite[Lynden-Bell 1967]{lyn67}, 
\cite[Baumgardt \etal\ 2008]{bau08}) but GCs are dynamically
evolved systems, hence mass loss is predicted during all their lifetime. It is
then possible that many clusters dissolved, and what we see now are the smaller
survivors of a potentially much larger population.

\begin{figure}[b]
% \vspace*{-2.0 cm}
\begin{center}
 \includegraphics[width=3.4in]{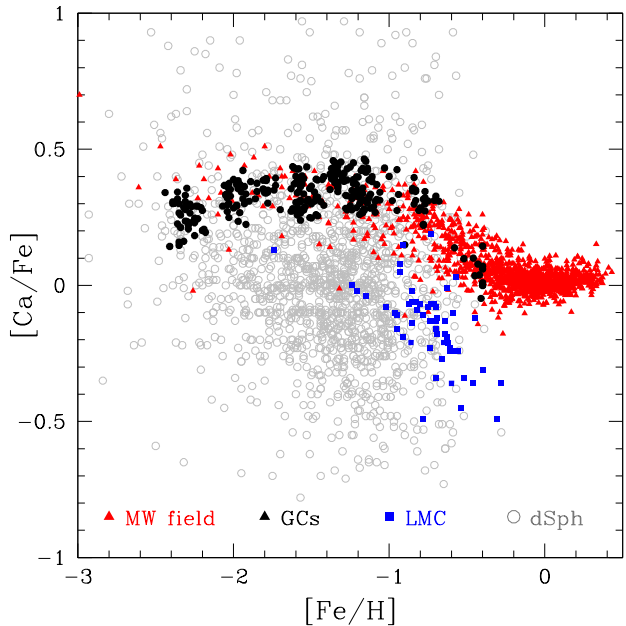} 
% \vspace*{-1.0 cm}
 \caption{[Ca/Fe] as a function of [Fe/H] for several stellar populations. Stars
 in GCs
 (black filled circles) from \cite[Carretta \etal\ (2010a)]{car10a}, 
 \cite[Carretta \etal\ (2010c)]{car10c},
 \cite[Carretta \etal\ (2011)]{car11},
 \cite[Carretta \etal\ (2013)]{car13},
 \cite[Carretta \etal\ (2014a)]{car14a},
 \cite[Carretta \etal\ (2014b)]{car14b},
 \cite[Carretta \etal\ (2015)]{carr15}; in dwarf galaxies (open grey circles) 
 from 
 \cite[Kirby \etal\ (2011)]{kir11}; in LMC (filled squares) from 
 \cite[Pompeia \etal\ (2008)]{pom08}; and in the Milky Way (filled red triangles) from
 \cite[Adibekyan \etal\ (2012)]{adi12}, 
 \cite[Chen \etal\ (2000)]{che00},
 \cite[Gratton \etal\ (2003)]{gra03}, and
 \cite[Jonsell \etal\ (2005)]{jon05}.}
   \label{fig1}
\end{center}
\end{figure}

Early (e.g. \cite[Ibata \etal\ 1994]{iba94}, \cite[Fusi Pecci \etal\ 1995]{fus95})
and more recent studies (\cite[Forbes and Bridges 2010]{for10}, 
\cite[Leaman \etal\ 2013]{lea13})
 on the accretion history focussed
on dwarf galaxies and entire clusters or entire systems of GCs, not
on the direct contribution from the clusters themselves. A possible reason is summarized
in Fig.\,\ref{fig1}: the chemical tagging made with $\alpha-$elements may 
distinguish between clusters and dwarf galaxies, but usually GC stars are 
superimposed to several Galactic components, formed in situ, accreted or even 
kicked out (see \cite[Sheffield \etal\ 2012]{she12}). 
Clearly we need another diagnostic.

\section{The Na-O anticorrelation}

\begin{figure}[b]
% \vspace*{-2.0 cm}
\begin{center}
 \includegraphics[width=3.4in]{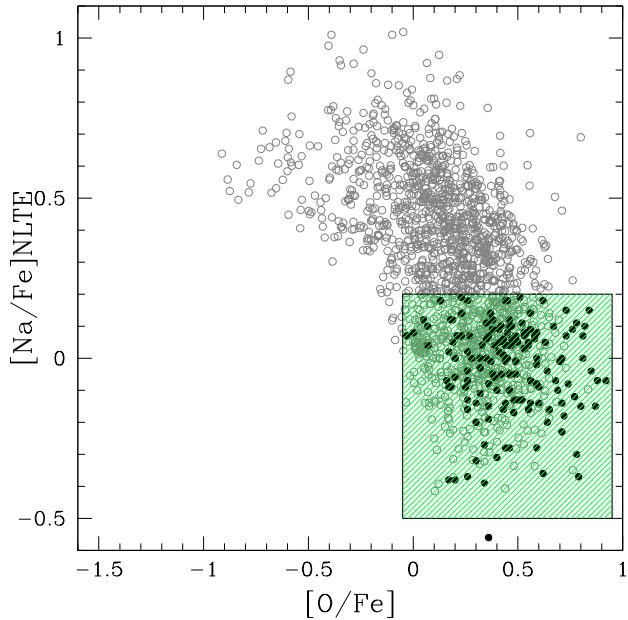} 
% \vspace*{-1.0 cm}
 \caption{[Na/Fe] as a function of [O/Fe] for GC (empty circles, 
 \cite[Carretta \etal\ 2009]{car09} and references in Fig.1) and field 
 stars (filled circles, \cite[Gratton \etal\ 2003]{gra03} and 
 \cite[Gratton \etal\ 2000]{gra00}). The green shaded box encompasses the field
 stars.}
   \label{fig2}
\end{center}
\end{figure}

One of the best tools for chemical tagging of GC stars is the Na-O
anticorrelation, which represents the unique DNA of GCs (see Fig.~\ref{fig2}
which summarizes the results of our FLAMES survey with $\sim 2500$ red giant
stars in 25 GCs, see 
\cite[Carretta 2015]{car15} for
references). The huge spreads in Na and O tell us that GCs are not simple 
stellar populations (by definition, coeval stars with the same initial chemical
composition) because GC stars have very different Na, O contents. Hence, their
stars are not strictly coeval: huge chemical differences translate
into tiny age differences, $10^6-10^7$ yrs, depending on the kind of polluter
chosen to release enriched matter in the intracluster gas (see the review by
\cite[Gratton \etal\ 2012]{gra12} for details).

This chemistry is explained as the result of proton-capture
reactions in H-burning at high temperature, enhancing or destroying different
elements according to the temperature stratification 
(\cite[Denisenkov \& Denisenkova 1989]{den89}).

The occurrence of this pattern is only seen in the high density GC environment, whereas in
Galactic Pop. II field stars O and Na are basically
untouched from the main sequence up to the giant branch (shaded box in 
Fig.~\ref{fig2}, where the field sample is taken
from \cite[Gratton \etal\ 2003]{gra03}) and \cite[Gratton \etal\ (2000)]{gra00}.

The Na-O anticorrelation, discovered by the Lick-Texas group 
(e.g. \cite[Kraft 1994]{kra94}),
joined other features like the anticorrelation of C and N, observed even in
unevolved stars, and of other elements along increasing Coulomb barrier like Al,
Mg and even K (\cite[Mucciarelli \etal\ 2012]{muc12}, 
\cite[Cohen \& Kirby 2012]{coh12}, \cite[Carretta 2014]{car14}). 
In turn, these abundance variations have an impact on the photometric
sequences in the color-magnitude diagrams due to molecular bands, mostly of
CNO elements 
(\cite[Sbordone \etal\ 2011]{sbo11}, \cite[Milone \etal\ 2012]{mil12}, 
\cite[Larsen \etal\ 2014a]{lar14a}). 
For heavier species there is no direct photometric observable.

The anticorrelation of Na and O is the most notable of these features because  
it is observed
in GCs spanning the whole mass range from the tiny Pal~5 
(\cite[Smith \etal\ 2002]{smi02}) in
dissolution phase to $\omega$ Cen 
(e.g. \cite[Johnson \& Pilachowski 2010]{joh10}), the
likely remnant of a nucleated dwarf galaxy; it is observed in GCs of sure or 
suspected extragalactic origin 
(e.g. M~54, \cite[Carretta \etal\ 2010a]{car10a}) and in those
showing a metallicity spread ($\omega$ Cen, M~54). This signature does  not
discriminate between GCs formed in situ or  likely accreted by our Galaxy
(\cite[Leaman \etal\ 2013]{lea13}). In short, it is a widespread feature among GCs, likely
related to the intrinsic mechanism of formation: we conclude that \textit{ a genuine old
GC is a system that formed in a short time at least two generations of stars with
different content of proton-capture elements}.

This signature allows us to put strong constraints on the initial masses of GCs.
An almost constant fraction ($\sim 33\%$) of first generation (FG) stars with 
primordial  composition still resides in GCs 
(\cite[Carretta \etal\ 2009]{car09}). However the
bulk of present day GC stars is composed of second generation (SG)  stars, with
modified composition. Therefore,  since SG stars are formed using the ejecta of
only a fraction of FG stars, and the ejecta from the main candidate polluters
are not enough (\cite[de Mink \etal\ 2009]{dem09}), we incur in the so 
called {\textit mass budget problem}. This is  solved with either a peculiar 
IMF or assuming that the
precursors of GCs were much more massive than present-day GCs, losing about 90\%
of their stars and becoming good candidates as main contributors to the halo.

\section{Multiple populations in GCs: second generation stars in the halo}

A selective mass loss is expected because SG stars form more centrally
concentrated in many models (e.g. \cite[D'Ercole \etal\ 2008]{der08}, 
\cite[Decressin \etal\ 2007]{dec07})
and also the following dynamical evolution will push
FG stars more easily outside tidal radius.
Observations show that currently SG stars are usually more concentrated (e.g. 
\cite[Milone \etal\ 2012]{mil12}, \cite[Carretta 2015]{car15}); however, after 
reaching full dynamical 
mixing, also SG stars with their unique chemical signature may be lost from 
GCs. These stars provide a clearcut probe of the contribution of GCs to the halo.

\cite[Carretta \etal\ (2010b)]{car10b} made a first attempt by comparing sodium abundances in
field stars and SG stars in GCs. They found a small fraction (1.4\%) of stars 
with GC signature evaporated in the halo (the fraction is doubled by 
considering the current ratio of FG/SG).

A more systematic survey found a similar fraction of SG stars by looking for
large N excess in metal-poor halo stars with spectra from the Sloan survey.
After cleaning the sample for contamination by thick disk stars 
\cite[Martell \etal\ (2011)]{mar11} found 3\% of halo stars with SG composition, 
in agreement with serendipitous discoveries using O and Na 
(\cite[Ramirez \etal\ 2012]{ram12}), unless one of their two O-poor stars is 
found to be polluted by mass
transfer from a companion AGB star, as suggested by the large abundance of 
barium and yttrium.

Finally, signatures of SG stars (Na, Al enhancements and O, Mg depletions) are 
more and more used to retrieve a cluster origin for streams 
(e.g. \cite[Sesar \etal\ 2015]{ses15}: Ophiuchus stream; 
\cite[Wylie-de Boer \etal\ 2012]{wyl12}: Aquarius stream)

Using these signatures many studies 
(\cite[Carretta \etal\ 2010b]{car10b}, 
\cite[Vesperini \etal\ 2010]{ves10}, 
\cite[Schaerer \& Charbonnel 2011]{sch11}, 
\cite[Martell \etal\ 2011]{mar11}, 
\cite[Gratton \etal\ 2012]{gra12})
provide estimates for the fraction of halo stars originated in GCs. These range 
from a few percent, using only observations, up to
almost 50\% of the halo mass, if the initial mass of GCs is assumed to be many 
times larger than present-day values.

\section{Multiple populations in GCs: first generation stars in the halo}

\begin{figure}[b]
% \vspace*{-2.0 cm}
\begin{center}
 \includegraphics[width=3.4in]{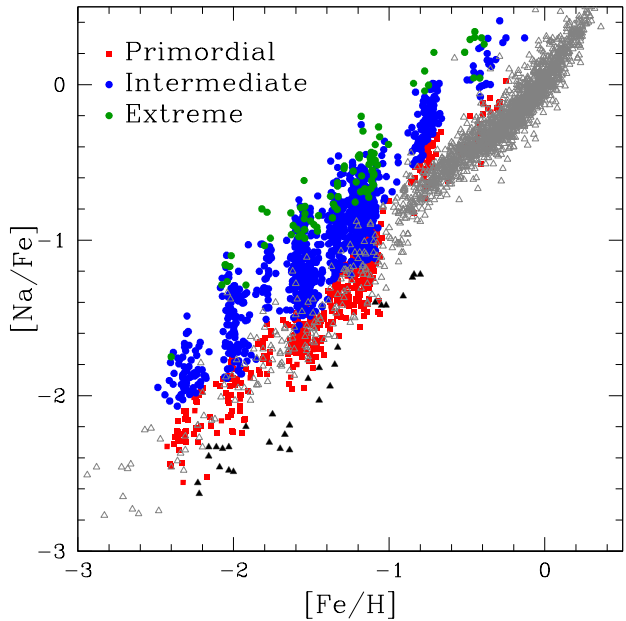} 
% \vspace*{-1.0 cm}
 \caption{[Na/H] as a function of [Fe/H] for GC stars of different generations 
 (filled squares and circles) and field stars (triangles). Filled triangles are
 the accreted field component.}
   \label{fig3}
\end{center}
\end{figure}

We get important information also from the FG component, which is much more
elusive, since FG stars share the same chemistry of field stars of similar
metallicity. In the homogeneous set
of Na abundance from Carretta (2013, empty triangles in Fig.3 here) there are a few stars clearly below the
general trend (filled triangles in Fig.~\ref{fig3}): these are objects 
tagged as accreted stars 
(e.g. \cite[Gratton \etal\ 2003]{gra03}, \cite[Nissen \& Schuster 2010]{nis10})
and their abundances are clearly similar to those observed in dwarf satellites of
Milky Way.
Therefore, we may infer that a minority of halo stars has a composition similar to the
dwarfs, but the bulk of halo stars looks like the FG component of GCs
(red filled squares in Fig.~\ref{fig3}).

This confirms the findings from $\alpha-$elements 
(\cite[Tolstoy \etal\ 2009]{tol09}) and 
carbon (\cite[Kirby \etal\ 2015b]{kir15b}): the majority of field halo stars cannot have formed
in present-day dSphs.

\begin{figure}[b]
% \vspace*{-2.0 cm}
\begin{center}
 \includegraphics[width=3.4in]{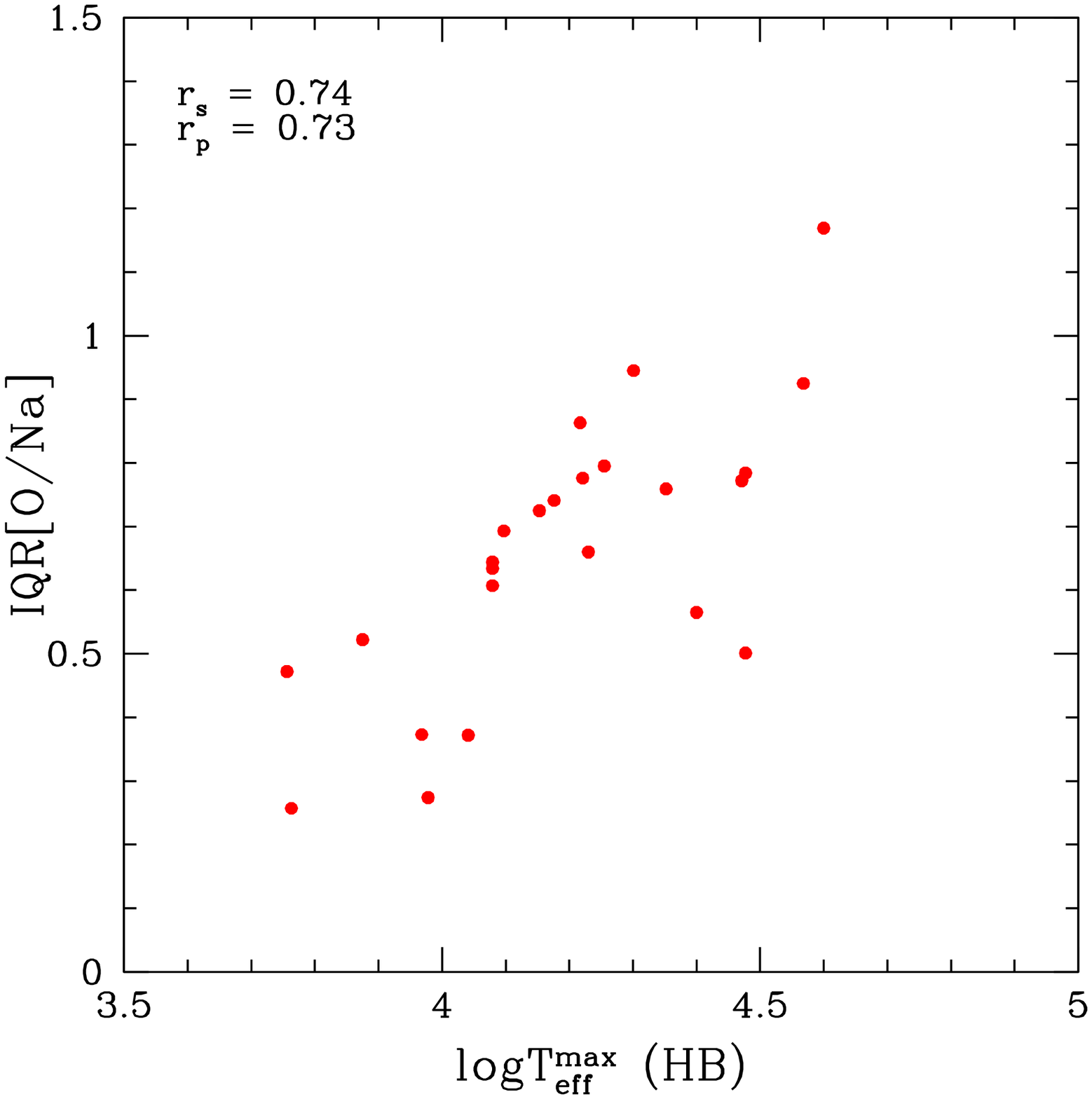} 
% \vspace*{-1.0 cm}
 \caption{The extension of the Na-O anticorrelation (measured by the
 interquartile range IQR[O/Na] as a function of the maximum temperature along
 the HB for GCs in the FLAMES survey 
 (\cite[Carretta \etal\ 2007]{car07}, \cite[Carretta \etal\ 2010b]{car10b}).}
   \label{fig4}
\end{center}
\end{figure}

Another supporting evidence comes from the horizontal branch (HB) morphology. We
know that the luminosity function of BHB stars is very different in GCs and in
the halo 
(e.g. \cite[Kinman \& Allen 1996]{kin96}, \cite[Gratton \etal\ 2012]{gra12}). 
We further know that 
Na-rich stars are also more enriched in He. 
\cite[Carretta \etal\ (2007)]{car07}, \cite[Carretta \etal\ (2010b)]{car10b}  
found
that the extension of the Na-O anticorrelation is correlated to the maximum
temperature reached at the bluest HB region
(see Fig.~\ref{fig4}) through the He abundance, increasing blueward.
The maximum temperature on the HB of field BHB stars implies that they are not
He enhanced. Since He-rich, Na-rich stars are rare in the field this suggests
that the field BHB are related to the FG in GCs.

Additional evidence comes from binaries: \cite[Goodwin (2010)]{god10} summarizes how the 
frequency and separation of binaries in  different environments can be used to
estimate the past density of their birth place. 
\cite[D'Orazi \etal\ (2010)]{dor10} and \cite[Lucatello \etal\ (2015)]{luc15} 
found that
the binary fraction in large samples  of GC stars is much lower among SG
stars than among FG stars, with evidence of a formation in much denser environment, as predicted by
theoretical models (\cite[Hong \etal\ 2015]{hon15}), and suggesting a common origin for the
field stars and  the FG populations.

\section{Challenges and conclusions}

The basic assumptions for a large contribution of GC stars to the halo 
(initial masses much larger, with selective loss of FG stars) must
confront with two major challenges.

Evidence is accumulating that in dwarf galaxies GCs cannot have been initially 
more than 4-5 times today's mass, as their present-day masses account for about 25\% of the galaxy
mass in metal poor stars 
(\cite[Larsen \etal\ 2012]{lar12}, \cite[Larsen \etal\ 2014b]{lar14b}, 
\cite[Tudorica \etal\ 2015]{tud15}). However, GCs
like those in the Fornax dSph show normal evidence of multiple populations 
(\cite[Larsen \etal\ 2014a]{lar14a}).

Furthermore, simulations of mass loss from gas expulsion 
(\cite[Khalaj \& Baumgardt 2015]{kha15}) predict a strong anticorrelation between the fraction of SG stars and the
cluster mass, that is not observed. Moreover,  no variation of the fraction of
SG stars as a function of cluster mass or  Galactocentric distance or
metallicity is observed (\cite[Bastian \& Lardo 2015]{bas15}), at odds with what expected by
mechanisms of mass loss.

The above points directly concern the multiple population scenario. However, 
independently from this scenario, larger birth cluster masses and
large amount of mass loss are predicted, providing a mass comparable
with the halo total mass 
(e.g. \cite[Marks \& Kroupa 2010]{mar10}, \cite[Fall \& Zhang 2001]{fal01}). 
In particular, primordial residual-gas expulsion and infant mortality may
account for the bulk of the halo with FG signature 
(\cite[Baumgardt \etal\ 2008]{bau08}).

In summary, we have plenty of evidence for a relevant contribution of
GCs to the halo: streams likely originated from GCs (from their size or chemical 
tagging, see also Grillmair, this volume), tidal tails, direct or indirect 
evidence from chemistry, HB stars and binary fraction.
Considering that a good third of the halo seems due to the so-called ``big
four" accretion events (Sagittarius, Hercules-Aquila, GASS, Virgo Cloud; see
\cite[Belokurov 2013]{bel13}), a large contribution of GCs to the general
assembly of the Galactic halo seems to be a viable option.


\begin{thebibliography}{}
\bibitem[Adibekyan \etal\ (2012)]{adi12}
{Adibekyan, V. Zh, Sousa, S.G., Santos, N.C., Delgado Mena, E.,
  Gonzales Hernandez, J.I., Israelian, G., Mayor, M., Khachatryan, G.} 2012,
\textit{A\&A}, 545, A32

\bibitem[Bastian \& Lardo (2015)]{bas15}
{Bastian, N., Lardo, C.} 2015,
\textit{MNRAS}, 453, 357

\bibitem[Baumgardt \etal\ (2008)]{bau08}
{Baumgardt, H., Kroupa, P., Parmentier, G.} 2008,
\textit{MNRAS}, 384, 1231

\bibitem[Belokurov (2013)]{bel13}
{Belokurov, V.} 2013,
\textit{New AR}, 57, 100

\bibitem[Bonifacio \etal\ (2015)]{bon15}
{Bonifacio, P., Caffau, E., Zaggia, S., Fran\c ois, P., Sbordone, L.,
Andrievsky, S.M., Korotin, S.A.} 2015,
\textit{A\&A}, 579, L6

\bibitem[Carretta (2013)]{car13}
{Carretta, E.} 2013,
\textit{A\&A}, 557, A128

\bibitem[Carretta (2014)]{car14}
{Carretta, E.} 2014,
\textit{ApJ}, 795, L28

\bibitem[Carretta (2015)]{car15}
{Carretta, E.} 2015,
\textit{ApJ}, 810, 148

\bibitem[Carretta \etal\ (2007)]{car07}
{Carretta, E., Recio-Blanco, A., Gratton, R.G., Piotto, G.,
Bragaglia, A.} 2007,
\textit{ApJ}, 671, L125

\bibitem[Carretta \etal\. (2009)]{car09}
{Carretta, E., Bragaglia, A., Gratton, R.G., et al.} 2009,
\textit{A\&A}, 505, 117 

\bibitem[Carretta \etal\ (2010a)]{car10a}
{Carretta, E., Bragaglia, A., Gratton, R.G., et al.} 2010a,
\textit{A\&A}, 520, 95

\bibitem[Carretta \etal\ (2010b)]{car10b}
{Carretta, E., Bragaglia, A., Gratton, R.G., Recio-Blanco, A.,
Lucatello, S., D'Orazi, V., Cassisi, S.} 2010b,
\textit{A\&A}, 516, 55

\bibitem[Carretta \etal\ (2010c)]{car10c}
{Carretta, E., Bragaglia, A., Gratton, R., Lucatello, S.,
  Bellazzini, M., D'Orazi, V.} 2010c,
\textit{ApJ}, 712, L21

\bibitem[Carretta \etal\ (2011)]{car11}
{Carretta, E., Lucatello, S., Gratton, R.G., Bragaglia, A., D'Orazi,
  V.} 2011,
\textit{A\&A}, 533, 69

\bibitem[Carretta \etal\ (2013)]{car13}
{Carretta, E., Bragaglia, A., Gratton, R.G. et al.} 2013,
\textit{A\&A}, 557, A138

\bibitem[Carretta \etal\ (2014a)]{car14a}
{Carretta, E., Bragaglia, A., Gratton, R.G., D'Orazi, V., Lucatello,
  S., Sollima, A.} 2014a,
\textit{A\&A}, 561, A87

\bibitem[Carretta \etal\ (2014b)]{car14b}
{Carretta, E., Bragaglia, A., Gratton, R.G. et al.} 2014b,
\textit{A\&A}, 564, A60

\bibitem[Carretta \etal\ (2015)]{carr15}
{Carretta, E., Bragaglia, A., Gratton, R.G. et al.} 2015,
\textit{A\&A}, 578, A116

\bibitem[Chen \etal\ (2000)]{che00}
{Chen, Y.Q., Nissen, P.E., Zhao, G., Zhang, H.W., Benoni, T.} 2000,
\textit{A\&AS}, 141, 491

\bibitem[Cohen \& Kirby (2012)]{coh12}
{Cohen, J.G., Kirby, E.N.} 2012,
\textit{ApJ}, 760, 86

\bibitem[Decressin \etal\ (2007)]{dec07}
{Decressin, T., Meynet, G., Charbonnel C. Prantzos, N.,
 Ekstrom, S.} 2007,
\textit{A\&A}, 464, 1029

\bibitem[de Mink \etal\ (2009)]{dem09}
{de Mink, S.E., Pols, O.R., Langer, N., Izzard, R.G.} 2009,
\textit{A\&A}, 507, L1

\bibitem[Denisenkov and Denisenkova (1989)]{den89}
{Denisenkov, P.A., Denisenkova, S.N.} 1989,
\textit{A.Tsir.}, 1538, 11

\bibitem[D'Ercole \etal\ (2008)]{der08}
{D'Ercole, A., Vesperini, E., D'Antona, F., McMillan, S.L.W.,
Recchi, S.} 2008,
\textit{MNRAS}, 391, 825

\bibitem[D'Orazi \etal\ (2010)]{dor10}
{D'Orazi, V., Gratton, R.G., Lucatello, S., Carretta, E., Bragaglia,
  A., Marino, A.F.} 2010,
\textit{ApJ}, 719, L213

\bibitem[Fall and Zhang (2001)]{fal01}
{Fall, S.M., Zhang, Q.} 2001,
\textit{ApJ}, 561, 751

\bibitem[Forbes and Bridges (2010)]{for10}
{Forbes, D.A., Bridges, T.} 2010,
\textit{MNRAS}, 404, 1203

\bibitem[Fusi Pecci \etal\ (1995)]{fus95}
{Fusi Pecci, F., Bellazzini, M., Cacciari, C., Ferraro F.R.} 1995,
\textit{AJ}, 110, 1664

\bibitem[Goodwin (2010)]{god10}
{Goodwin, S.P.} 2010,
\textit{RSPTA}, 368, 851

\bibitem[Gratton \etal\ (2000)]{gra00}
{Gratton, R.G., Sneden, C., Carretta, E., Bragaglia, A.} 2000,
\textit{A\&A}, 354, 169

\bibitem[Gratton \etal\ (2003)]{gra03}
{Gratton, R.G., Carretta, E., Claudi, R., Lucatello, S., \&
  Barbieri, M.} 2003,
\textit{A\&A}, 404, 187

\bibitem[Gratton \etal\ (2012)]{gra12}
{Gratton, R.G., Carretta, E., Bragaglia, A.} 2012,
\textit{A\&AR}, 20, 50

\bibitem[Hong \etal\ (2015)]{hon15}
{Hong, J., Vesperini, E., Sollima, A., McMillan, S.L.W., D'Antona,
  F., D'Ercole, A.} 2015,
\textit{MNRAS}, 449, 629

\bibitem[Ibata \etal\ (1994)]{iba94}
{Ibata, R.A., Irwin, M.J., Gilmore, G.} 1994,
\textit{Nature}, 370, 194

\bibitem[Johnson \& Pilachowski (2010]{joh10}
{Johnson, C.I., Pilachowski, C.A.} 2010,
\textit{ApJ}, 722, 1373

\bibitem[Jonsell \etal\ (2005)]{jon05}
{Jonsell, K., Edvardsson, B., Gustafsson, B., Magain, P., Nissen,
  P.E., Asplund, M.} 2005,
\textit{A\&A}, 440, 321 

\bibitem[Khalaj \& Baumgardt (2015)]{kha15}
{Khalaj, P., Baumgardt, H.} 2015,
\textit{MNRAS}, 452, 924

\bibitem[Kinman \& Allen (1996)]{kin96}
{Kinman, T.D., Allen, C.} 1996,
\textit{ASP-CS}, 92, 36

\bibitem[Kirby \etal\ (2011)]{kir11}
{Kirby, E.N., Cohen, J.G, Smith, G.H., Majewski, S.R., Sohn, S.T., Guhathakurta,
P.} 2011,
\textit{ApJ}, 727, 79

\bibitem[Kirby \etal\ (2015a)]{kir15a}
{Kirby, E.N., Simon, J.D., Cohen, J.G.} 2015,
\textit{ApJ}, 810, 56

\bibitem[Kirby \etal\ (2015b)]{kir15b}
{Kirby, E.N., Guo, M., Zhang, A.J., et al.} 2015,
\textit{ApJ}, 801. 125

\bibitem[Kraft (1994)]{kra94}
{Kraft, R.P.} 1994,
\textit{PASP}, 106, 553

\bibitem[Larsen \etal (2012)]{lar12}
{Larsen, S.S., Strader, J., Brodie, J.P.} 2012,
\textit{A\&A}, 544, L14

\bibitem[Larsen \etal\ (2014a)]{lar14a}
{Larsen, S.S., Brodie, J.P., Forbes, D.A., Strader, J.} 2014a,
\textit{A\&A},565, A98 

\bibitem[Larsen \etal\ (2014b)]{lar14b}
{Larsen, S.S., Brodie, J.P., Grundahl, F., Strader, J.} 2014b,
\textit{ApJ}, 797, 15

\bibitem[Leaman \etal\ (2013)]{lea13}
{Leaman, R., VandenBerg, D.A., Mendel, J.T.} 2013,
\textit{MNRAS}, 436, 122

\bibitem[Lucatello \etal\ (subm.)]{luc15}
{Lucatello, S., Sollima, A., Gratton, R.G., Vesperini, E., D'Orazi, V.,
Carretta, E., Bragaglia, A.} 2015,
\textit{A\&A}, in press, arXiv:1509.05014

\bibitem[Lynden-Bell (1967)]{lyn67}
{Lynden-Bell, D.} 1967,
\textit{MNRAS}, 136, 101

\bibitem[Marks \& Kroupa (2010)]{mar10}
{Marks, M., Kroupa, P.} 2010,
\textit{MNRAS}, 406, 2000

\bibitem[Martell \etal\ (2011)]{mar11}
{Martell, S.L., Smolinski, J.P., Beers, T.C., Grebel, E.K.} 2011,
\textit{A\&A}, 534, 136

\bibitem[Milone \etal\ (2012)]{mil12}
{Milone, A., Piotto, G., Bedin, L. et al.} 2012,
\textit{ApJ}, 744, 58

\bibitem[Mucciarelli \etal\ (2012)]{muc12}
{Mucciarelli, A., Bellazzini, M., Ibata, R., Merle, T., Chapman,
S.C., Dalessandro, E., Sollima, A.} 2012,
\textit{MNRAS}, 426, 2889

\bibitem[Nissen \& Schuster (2010)]{nis10}
{Nissen, P.E., Schuster, W.J.} 2010,
\textit{A\&A}, 511, L10

\bibitem[Pompeia \etal\ (2008)]{pom08}
{Pomp\'eia, L., Hill, V., Spite, M., Cole, A. et al. 2008} 2008,
\textit{A\&A}, 480, 379

\bibitem[Ramirez \etal (2012)]{ram12}
{Ram\'irez, I., Mel\'endez, J., Chanam\'e, J.} 2012,
\textit{ApJ}, 757, 164

\bibitem[Sbordone \etal (2011)]{sbo11}
{Sbordone, L., Salaris, M., Weiss, A., Cassisi, S.} 2011,
\textit{A\&A}, 534, A9

\bibitem[Schaerer \& Charbonnel (2011)]{sch11}
{Schaerer, D., Charbonnel, C.} 2011,
\textit{MNRAS}, 413, 2297

\bibitem[Sesar \etal\ (2015)]{ses15}
{Sesar, B., Bovy, J., Bernard, E.J. et al.} 2015,
\textit{ApJ}, 809, 59

\bibitem[Sheffield \etal (2012)]{she12}
{Sheffield, A.A., Majewski, S.R., Johnston, K.V. et al.} 2012,
\textit{ApJ}, 761, 161

\bibitem[Smith \etal\ (2002)]{smi02}
{Smith, G.H., Sneden, C., Kraft, R.P.} 2002,
\textit{AJ}, 123, 1502

\bibitem[Tolstoy \etal\ (2009)]{tol09}
{Tolstoy, E., Hill, V., Tosi, M.} 2009,
\textit{ARA\&A}, 47, 371

\bibitem[Tudorica \etal\ (2015)]{tud15}
{Tudorica, A., Georgiev, I.Y., Chies-Santos, A.L.} 2015,
\textit{A\&A}, 581, 84

\bibitem[Vesperini \etal\ (2010)]{ves10}
{Vesperini, E., McMillan, S.L.W., D'Antona, F., D'Ercole, A.} 2010,
\textit{ApJ}, 718, 112

\bibitem[Wylie-de Boer \etal\ (2012)]{wyl12}
{Wylie-de Boer, E., Freeman, K., williams, M., Steinmetz, M.,
  Munari, U., Keller, S.} 2012,
\textit{ApJ}, 755, 35


\end{thebibliography}
\end{document}